\documentclass{jfm}
\usepackage{graphicx}
\usepackage[font=small,skip=0pt]{caption}
\usepackage{titlesec}
\titlespacing*{\section}{0pt}{1.1\baselineskip}{\baselineskip}
\usepackage{epstopdf, epsfig}
\usepackage[euler]{textgreek}
\usepackage{fixltx2e}
\usepackage{todonotes}
\usepackage{caption}
\usepackage{subfig}
\usepackage{tabularx}
\usepackage{wrapfig}
\usepackage{comment}
\usepackage{amsmath} 
\usepackage{cleveref}
\usepackage{amssymb}
\usepackage{amsfonts}
\usepackage{mathabx}
\usepackage{MnSymbol}
\usepackage{wasysym}

\usepackage{hhline}
\usepackage{multirow}
\usepackage{physics}
\usepackage{tabularx}
\usepackage{array}
\usepackage{subfig}
\usepackage{mathtools}
\usepackage{siunitx}
\newcolumntype{M}[1]{>{\centering\arraybackslash}m{#1}}
\usepackage{pdfpages}
\usepackage[running]{lineno}

\title{Neural Network aided quarantine control model estimation of COVID spread in Wuhan, China}
\author{ Raj Dandekar\aff{1} \and George Barbastathis\aff{2,3} \corresp{\email{gbarb@mit.edu}}}
\affiliation{\aff{1}Department of Civil and Environmental Engineering, Massachusetts Institute of Technology, Cambridge, MA 02139, USA, \aff{2} Department of Mechanical Engineering, Massachusetts Institute of Technology, Cambridge, MA 02139, USA, \aff{3}Singapore-MIT Alliance for Research and Technology (SMART) Centre, Singapore 138602} 

\begin{document}
\maketitle
In a move described as unprecedented in public health history, starting 24 January 2020, China imposed quarantine and isolation restrictions in Wuhan, a city of more than 10 million people. This raised the question: is mass quarantine and isolation effective as a social tool in addition to its scientific use as a medical tool? In an effort to address this question, using a epidemiological model driven approach augmented by machine learning, we show that the quarantine and isolation measures implemented in Wuhan brought down the effective reproduction number $R(t)$ of the CoVID-19 spread from $R(t) > 1$ to $R(t) <1$ within a month after the imposition of quarantine control measures in Wuhan, China. This ultimately resulted in a stagnation phase in the infected case count in Wuhan. Our results indicate that the  strict public health policies implemented in Wuhan may have played a crucial role in halting down the spread of infection and such measures should potentially be implemented in other highly affected countries such as South Korea, Italy and Iran to curtail spread of the disease. Finally, our forecasting results predict a stagnation in the quarantine control measures implemented in Wuhan towards the end of March 2020; this would lead to a subsequent stagnation in the effective reproduction number at $R(t) <1$. We warn that immediate relaxation of the quarantine measures in Wuhan may lead to a relapse in the infection spread and a subsequent increase in the effective reproduction number to $R(t) >1$. Thus, it may be wise to relax quarantine measures after sufficient time has elapsed, during which maximum of the quarantined/isolated individuals are recovered. 

(*\textit{It should be noted when we mention quarantine subsequently in the paper, we mean the infected people who are quarantined and isolated and cannot lead to an infection in a susceptible person.})
\section{Introduction}
 The Coronavirus respiratory disease 2019 originating from the virus ``SARS-CoV-2" \citep{chan2020familial, cdc} has led to a global pandemic, leading to $1,05,586$ confirmed global cases in $72$ countries as of March 8, 2020 \citep{worldcoronavirus}. Of these cases, $80, 859$ have been reported in Mainland China, followed by $7134$ cases in the Republic of Korea (Western Pacific Region), $5883$ cases in Italy (European Region) and $5823$ cases in Iran (Eastern Mediterranean Region) \citep{worldcoronavirus}. \newline
A quantitative measure to estimate the efficacy of a virus is its basic reproduction number, $R_{0}$ \citep{van2017reproduction}. $R_{0}$ represents the average number of infections generated by a single infected individual in a population of susceptible individuals. A number of studies focused on estimating $R_{0}$ for the COVID spread in China using initial data collected till 20 - 30 Jan 2020 \citep{imai2020report, read2020novel, tang2020estimation, li2020early, wu2020nowcasting}. Several estimates of $R_{0}$ made with different modelling approaches and assumptions were reported ranging from $2.6$ \citep{imai2020report}, $3.8$ \citep{read2020novel}, $6.47$ \citep{tang2020estimation}, $2.2$ \citep{li2020early} and $2.68$ \citep{wu2020nowcasting}, all of which predicted the number of infected cases in China to rise exponentially until preventive measures were rapidly undertaken.  However, most of the prior studies have analyzed the situation till $20 - 30$ January 2020, during which there was limited availability of temporal and spatially resolved data. As a result, many of these models parametrizations used information about prior outbreaks such as SARS to predict the temporal evolution of the disease spread and thus estimate the reproduction number. With detailed time and spatially resolved data available from the Chinese National Health Commission starting $20$ January 2020, there is an urgent need to use data driven approaches to estimate several factors governing the disease spread.  \newline
In a move described as unprecedented in public health history, starting at the end of January, China quarantined the entire Wuhan province housing over $10$ million people by shutting down its public transport system, train and airport stations. In addition to this, several strict public healthcare measures related to quarantine and isolation were imposed in Wuhan. In this study, in contrast to previous studies, we focus on the time just at the onset of quarantine control in Wuhan: $24$ January 2020 to 40 days after that: till $3$ March 2020; to quantify the effect of the imposed quarantine control in limiting the spread of CoVID-19. We define a neural network aided epidemiology model based on an effective reproduction number $R(t)$ which captures the quarantine control strength function $Q(t)$. Based on the time resolved data for the infected and recovered cases, the neural network is able to learn the increase of quarantine strength $Q(t)$ and the associated decrease of $R(t)$ with time; indicating the effect of quarantine control in preventing the infectious count from exponentially exploding in Wuhan. 
\begin{figure}
\centering
\begin{tabular}{cc}
\subfloat[]{\includegraphics[width=0.45\textwidth]{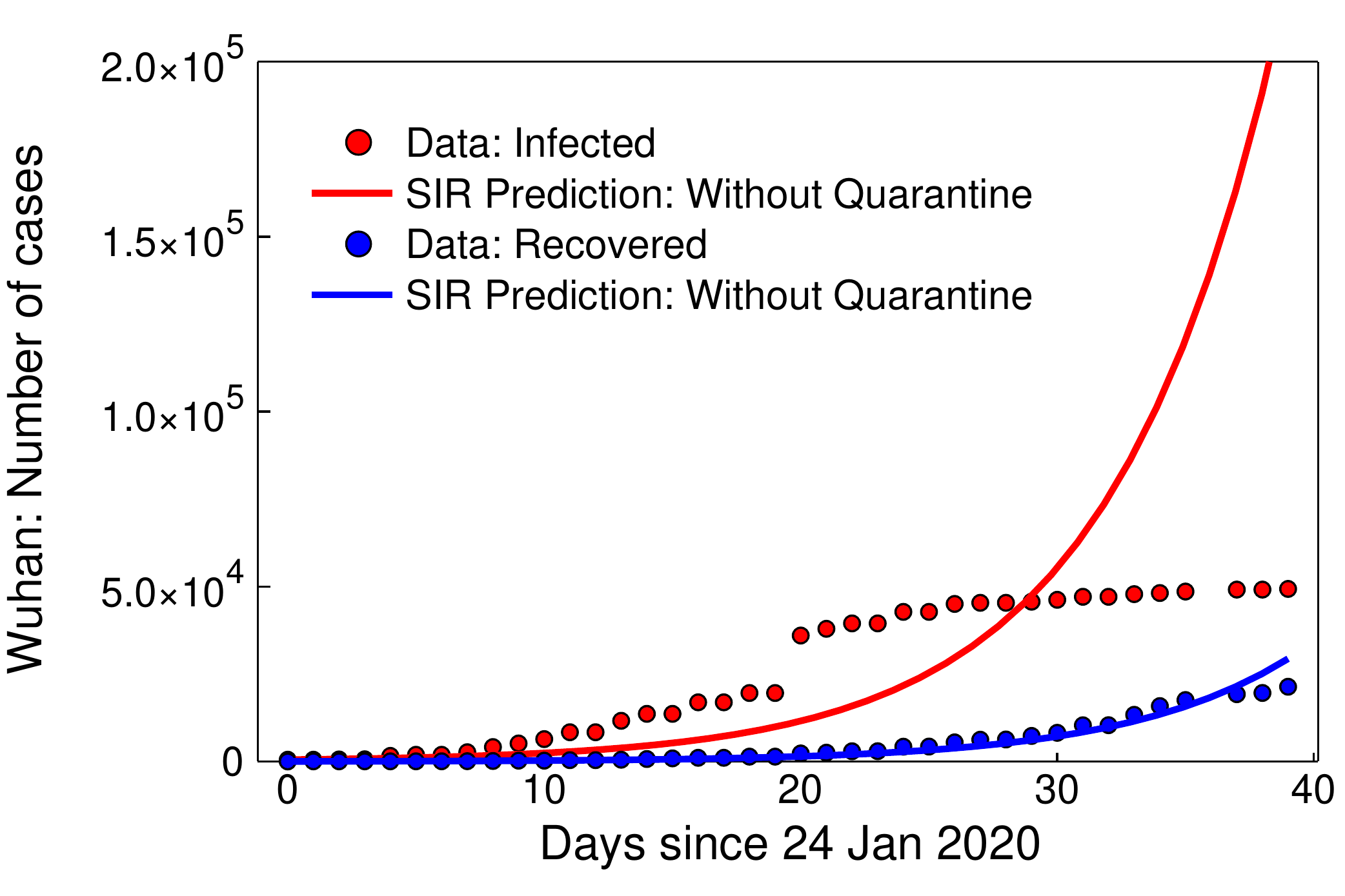}}
\subfloat[]{\includegraphics[width=0.45\textwidth]{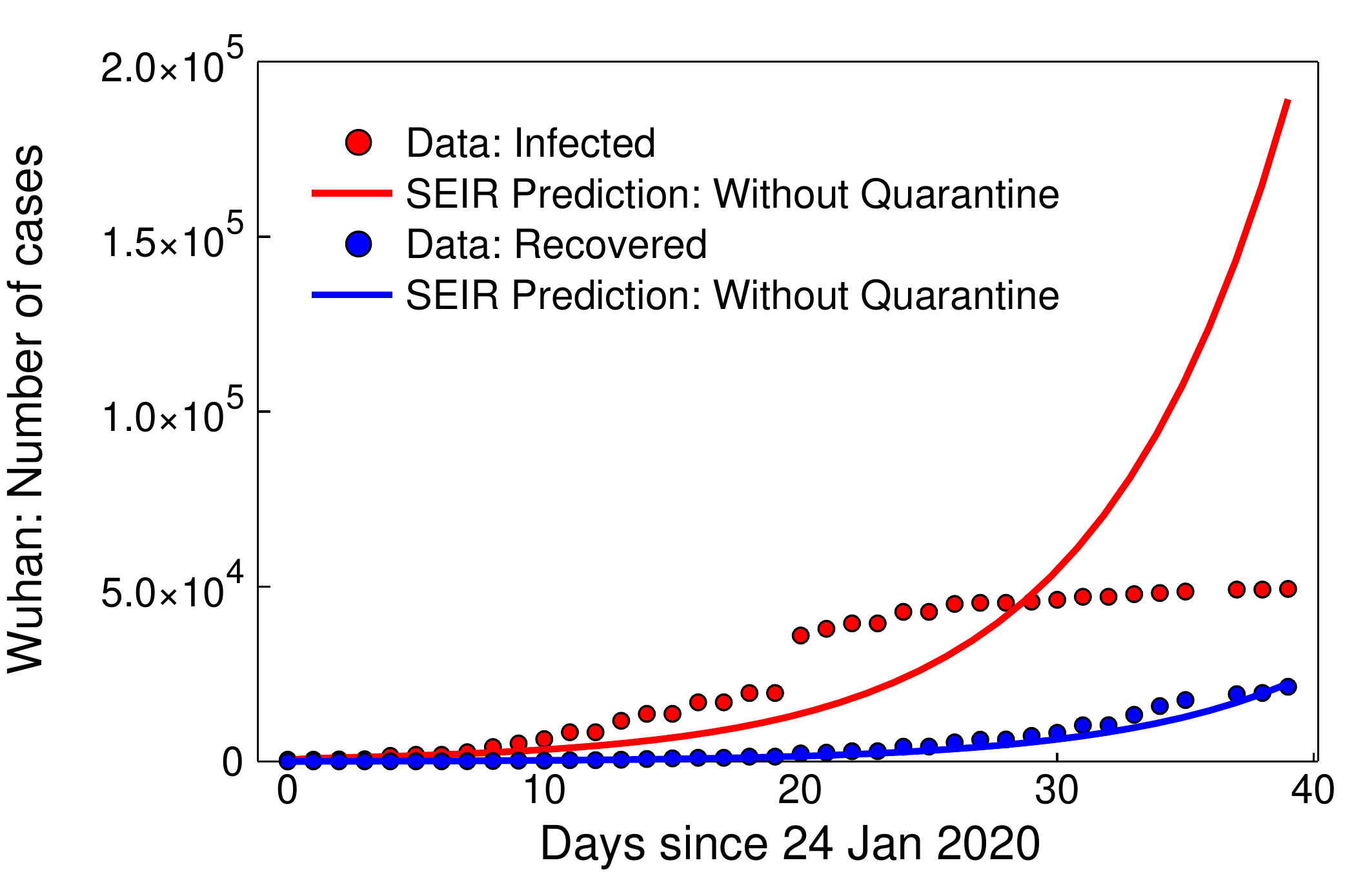}}\\
\end{tabular}
\caption{\textbf{Without quarantine control:} Estimation of the infected and recovered case count compared to the data acquired from the Chinese National Health Commission  post 24 January 2020 in Wuhan, China based on (a) SIR epidemiological model and (b) SEIR epidemiological model.}\label{figure1}
\end{figure}\label{figure1}

\begin{figure}
\centering
\begin{tabular}{cc}
\subfloat[]{\includegraphics[width=0.47\textwidth]{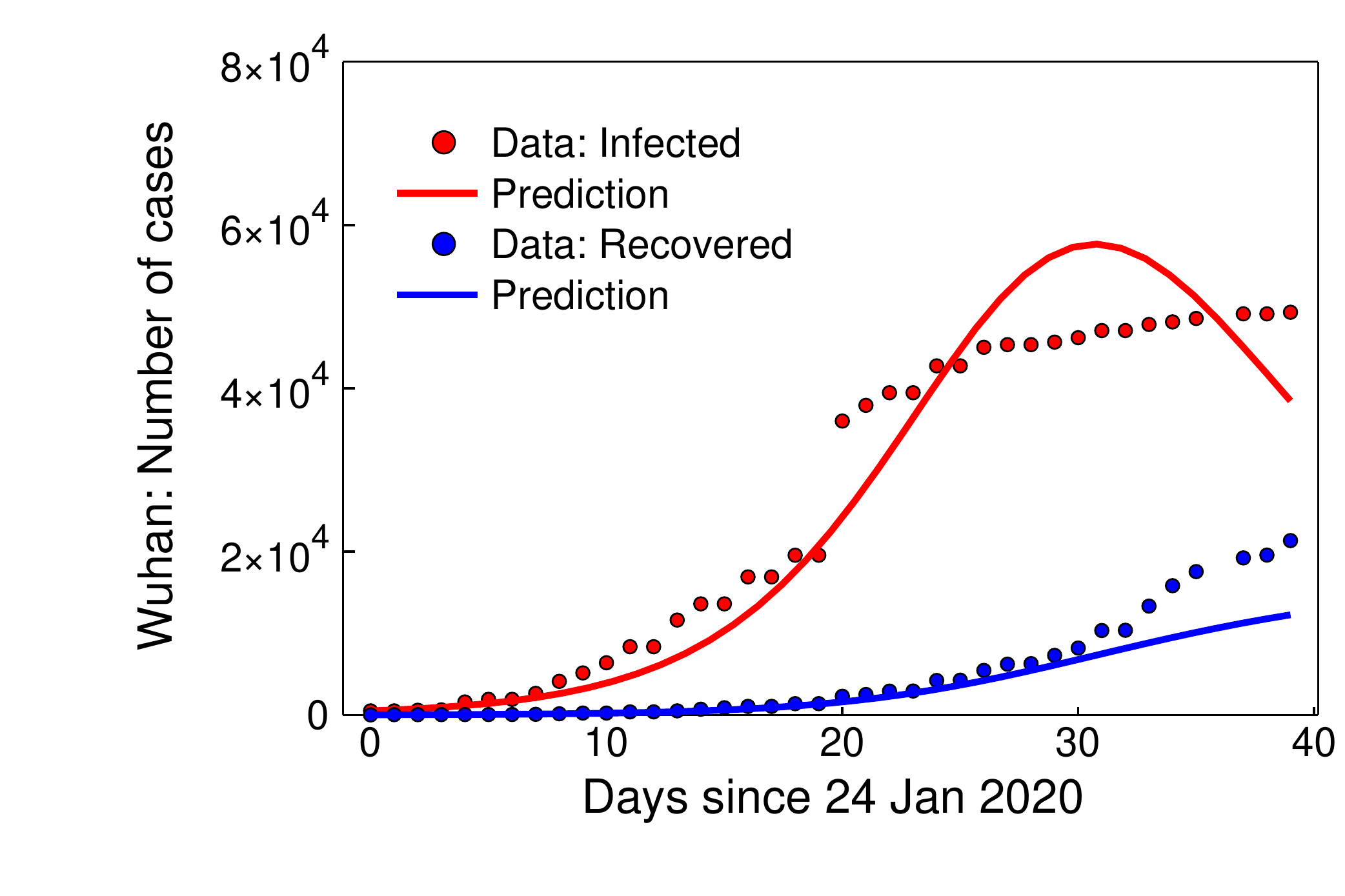}}
\subfloat[]{\includegraphics[width=0.47\textwidth]{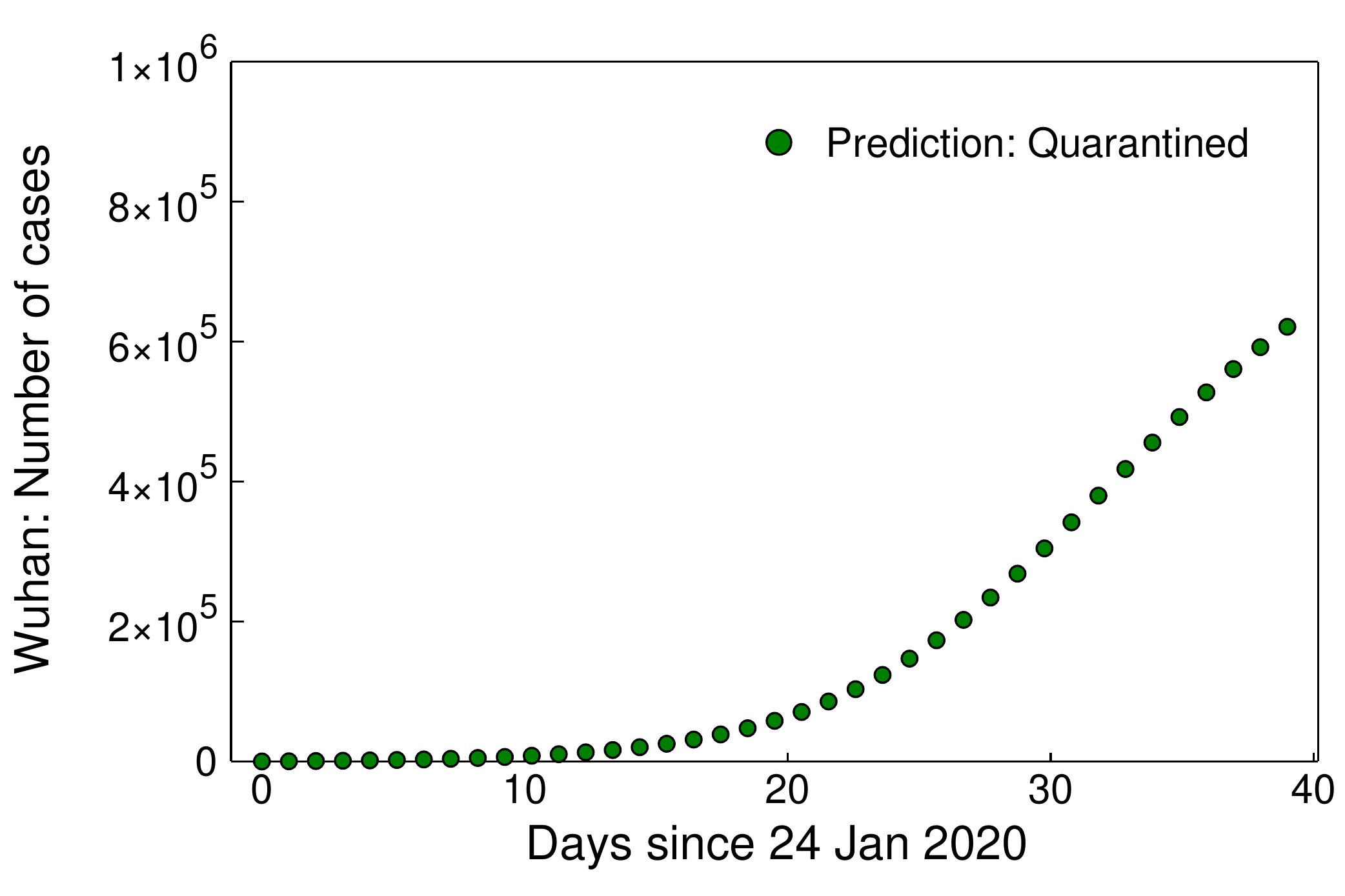}}\\
\subfloat[]{\includegraphics[width=0.47\textwidth]{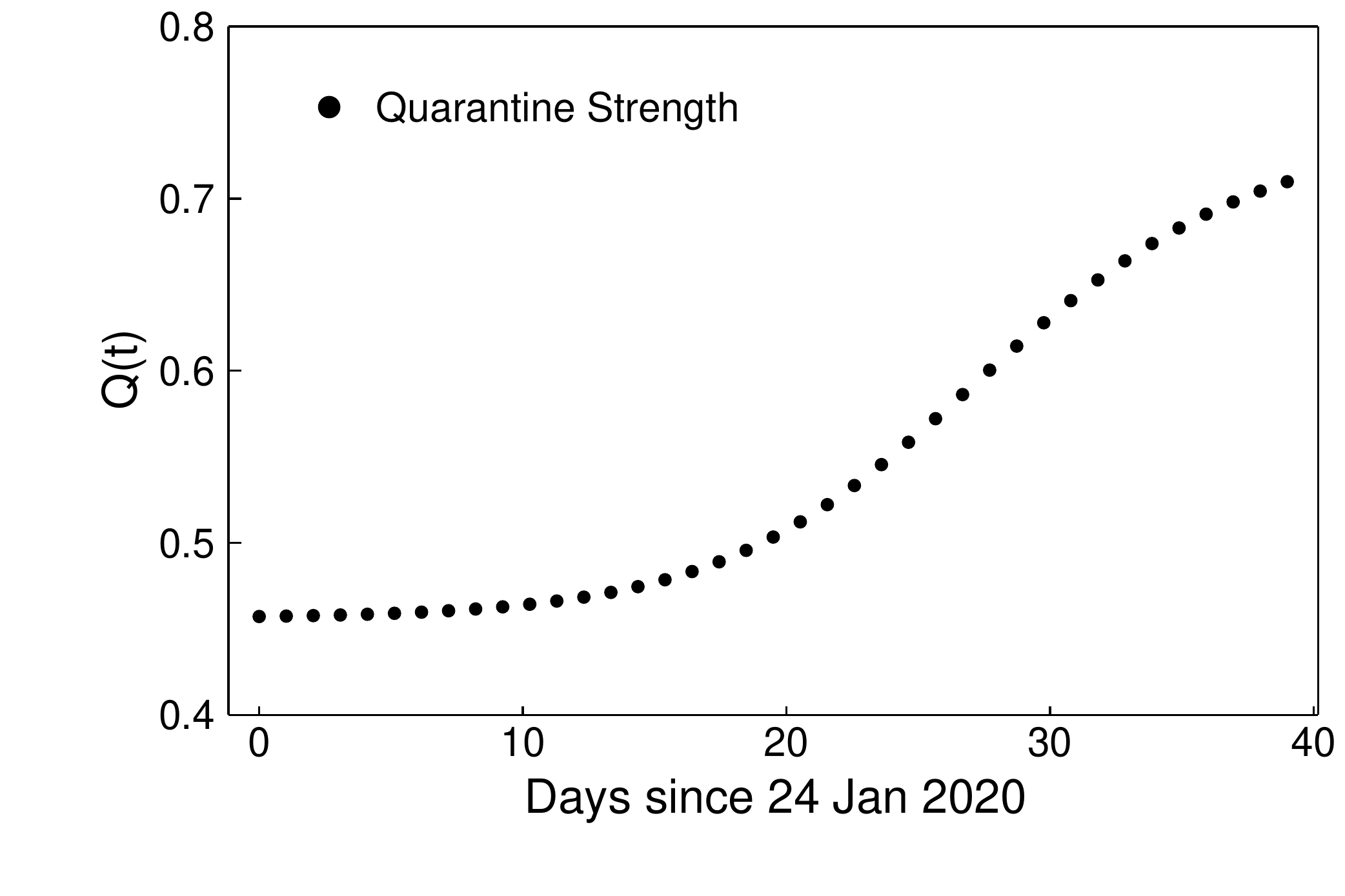}}
\subfloat[]{\includegraphics[width=0.47\textwidth]{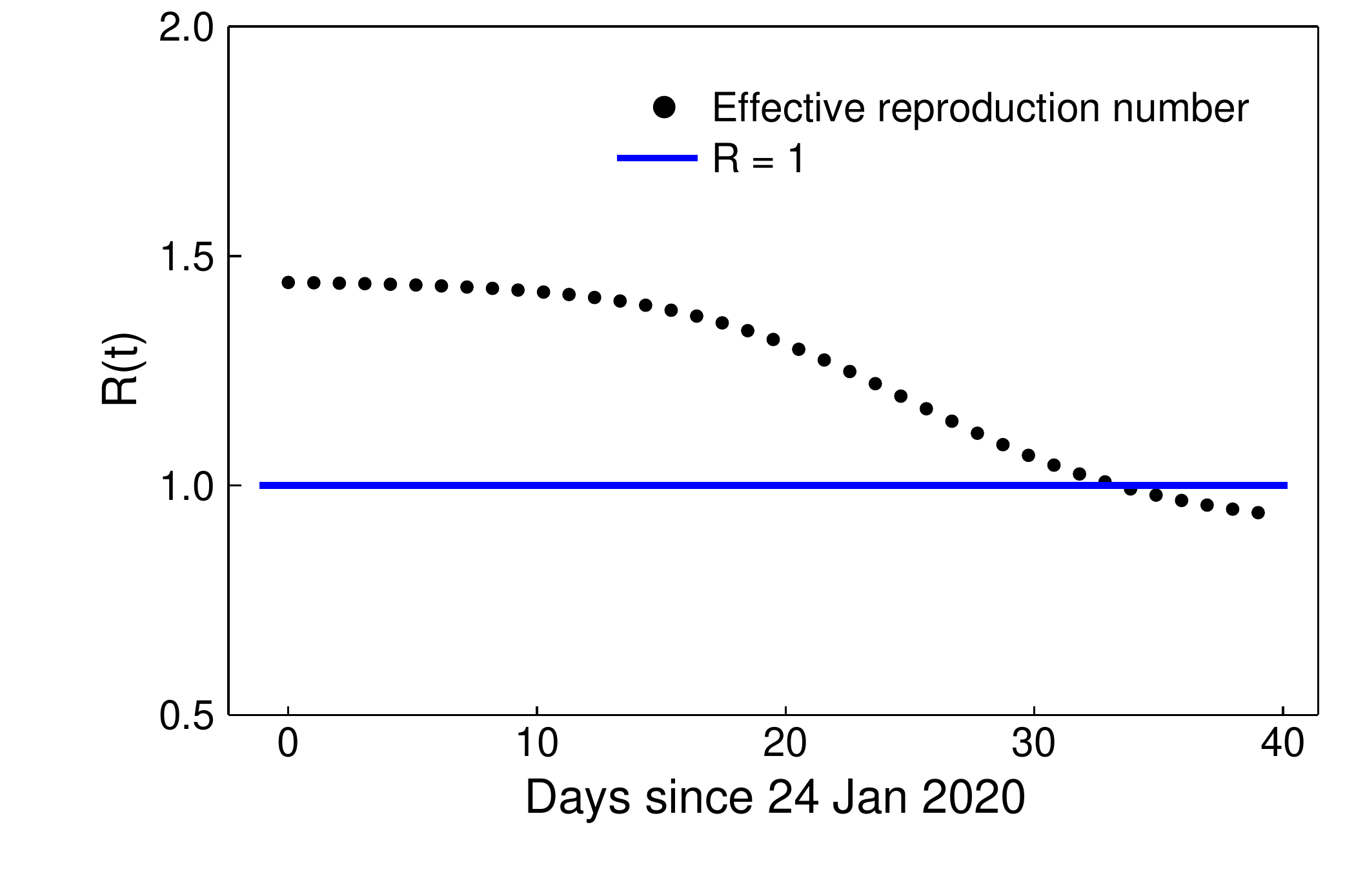}}
\end{tabular}
\caption{\textbf{Neural network aided SIR quarantine control model:} (a) Estimation of the infected and recovered case count compared to the data acquired from the Chinese National Health Commission post 24 January 2020 in Wuhan, China. (b) Number of quarantined people $T(t)$ with time. (c) Quarantine strength function $Q(t)$ learnt by the neural network. (d) Effective reproduction number $R(t)$ as function of time. }
\end{figure}\label{figure2}

\iffalse
\begin{figure}
\centering
\begin{tabular}{cc}
\subfloat[]{\includegraphics[width=0.47\textwidth]{RD_Nature_QSIR_1n.pdf}}\\
\subfloat[]{\includegraphics[width=0.45\textwidth]{RD_Nature_QSIR_2n.pdf}}
\subfloat[]{\includegraphics[width=0.45\textwidth]{RD_Nature_QSIR_3n.pdf}}
\end{tabular}
\caption{\textbf{Neural network aided SIR quarantine control model:} (a) Estimation of the infected and recovered case count compared to the data acquired from the Chinese National Health Commission post 24 January 2020 in Wuhan, China. (b) Quarantine strength function $Q(t)$ learnt by the neural network. (c) Effective reproduction number $R(t)$ as function of time. }
\end{figure}\label{figure2}
\fi

\iffalse
\begin{figure}
\centering
\begin{tabular}{ccc}
\subfloat[]{\includegraphics[width=0.32\textwidth]{RD_Nature_QSIR_10n.pdf}}
\subfloat[]{\includegraphics[width=0.307\textwidth]{RD_Nature_QSIR_11n.pdf}}
\subfloat[]{\includegraphics[width=0.315\textwidth]{RD_Nature_QSIR_12n.pdf}}

\end{tabular}
\caption{\textbf{$1$ month Forecasting of} (a) Quarantine strength, $Q$ and (b) Effective reproduction number, $R(t)$ in Wuhan, China based on the neural network augmented SIR model. }\label{figure3}
\end{figure}
\fi

\begin{figure}
\centering
\begin{tabular}{cc}
\subfloat[]{\includegraphics[width=0.45\textwidth]{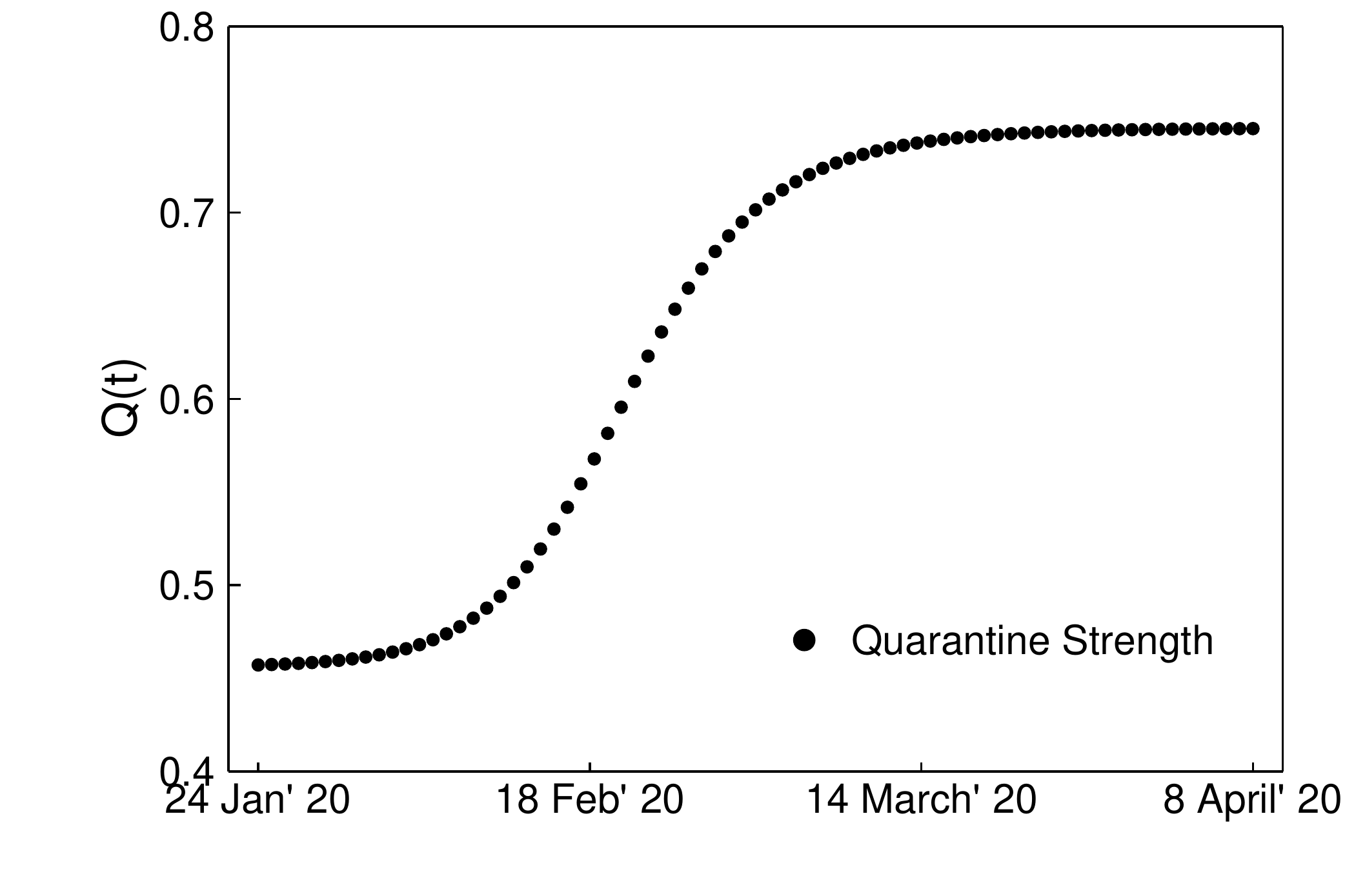}}
\subfloat[]{\includegraphics[width=0.45\textwidth]{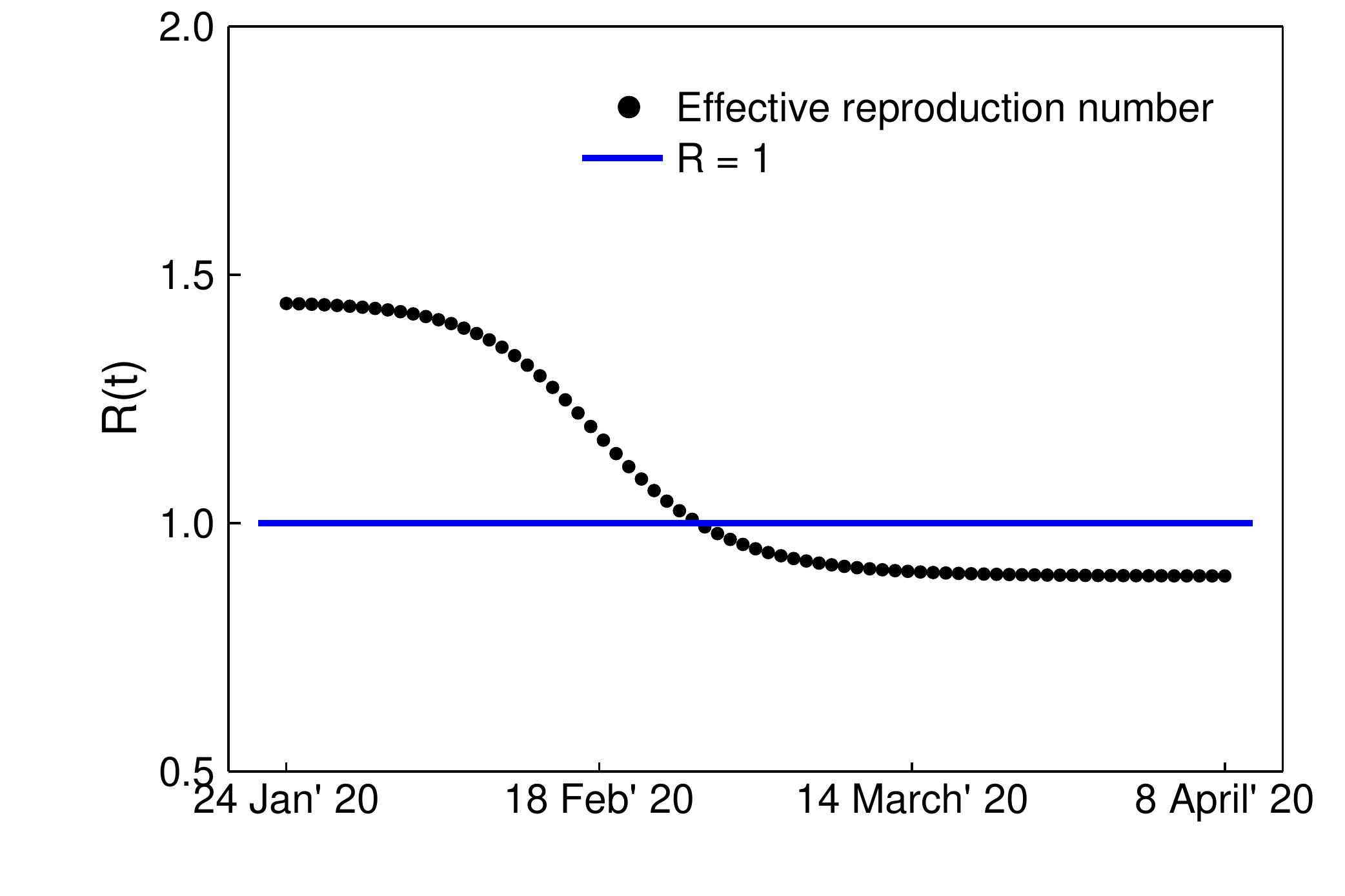}}

\end{tabular}
\caption{ \textbf{$1$ month Forecasting of} (a) Quarantine strength, $Q$ and (b) Effective reproduction number, $R(t)$ in Wuhan, China based on the neural network augmented SIR model. }\label{figure3}
\end{figure}

\section*{Results}
\textbf{Without quarantine control in Wuhan, China} \newline
If quarantine control is not considered, figure 1 shows that using both the classical SEIR and the SIR models with optimized model parameters, the number of infected cases continue to rise exponentially. Neither of these models can recover the stagnation seen in the infected number, about $30$ days post $24$ January 2020 when the quarantine measures  were implemented in Wuhan. Detailed calculations are shown in the Materials and Methods section. \newline
\textbf{With quarantine control in Wuhan, China} \newline
Figure 2a shows that a neural network augmented SIR model with quarantine control included,  captures well the time evolution of the infected and recovered case count, including the plateau value seen at about $30$ days post $24$ January 2020. Inclusion of quarantine control in the SIR model thus prevents the exponential growth in the infected case count seen in Figure 1. Figure 2b shows the model estimate of the number of quarantined/isolated people, which peaks at a value of $0.6$ million people. These individuals were effectively prevented from spreading infection through imposition of quarantine/isolation restrictions. Figure 2c shows the optimal quarantine strength function $Q(t)$ learned by the neural network,  representing the fraction of the infected people subjected to quarantine/isolation restrictions. Starting from about $0.5$ on $24$ January 2020, as the strength of the strict quarantine and isolation effects employed in Wuhan increased, this fraction increased to $0.7$ within a month, implying that a large fraction of the infected population were effectively prevented from transmitting the disease to the non-infected susceptible population. The quarantine strength function thus lies at the core of Wuhan's success in containing the infectious case count from exponentially increasing. The quarantine strength function is inversely related to the effective reproduction number of the virus. The effective reproduction number $R(t)$ based on the quarantine strength is plotted in Figure 2d, which shows a monotonic decrease with time. At the onset of quarantine control in Wuhan, i.e on $24$ January 2020; $R(t) \approx 1.5 > 1$ which is close to the value reported in prior studies \citep{imai2020report, read2020novel, tang2020estimation, li2020early, wu2020nowcasting}. Within a month after quarantine restrictions were imposed in Wuhan i.e around $23$ February 2020 , we see that $R(t) <1$, indicating the abatement in infection spread. \newline
$1$ month forecasting based on the neural network augmented SIR model is shown in figure 3. 
\iffalse
Our optimized model predicts a decrease in the infected case count in the month of March, with only a small fraction of Wuhan's population being likely to remain infected by mid April 2020. Our model estimates a steadily growing recovered case count. Following the decrease in the infected case count, the quarantine strength $Q(t)$ is predicted to stagnate as China relaxes quarantine measures in Wuhan, accompanied by a stagnation in the effective reproduction number at $R(t) <1$ indicating an abatement in the infection spread. Detailed calculations are shown in the Materials and Methods section.
\fi
Our optimized model predicts a stagnation of the quarantine strength at $Q(t) \approx 0.75$ as China eases quarantine measures in Wuhan, accompanied by a stagnation in the effective reproduction number at $R(t) <1$; indicating control of the epidemic. Detailed calculations are shown in the Materials and Methods section.
\section*{Discussion}
Since $24$ January 2020, all public transportation in and out of Wuhan was restricted, effectively locking down a city of about $11$ million people. Starting 9 February 2020, every building in urban areas and every village in rural area was quarantined in Wuhan to contain the spread of CoVID. In addition to this, unprecedented efforts were employed by the Chinese government to prevent human to human contact between susceptible and infected individuals. At the same time, the data for the infected case count in Wuhan was seen to enter a stagnation phase about $30$ days post $24$ January 2020, when quarantine measures were deployed in Wuhan. 
\iffalse
Several studies \cite{fraser2004factors, day2006quarantine} have considered whether quarantine and isolation are effective measures in curtailing the spread of an infectious disease. The answer depends on several factors such as the basic reproduction number, as well as the extent to which the disease is transmitted asymptomatically.
\fi
In this study, we show that the classical SIER and the SIR models in which the model parameters, {\it i.e.} the transmission rate $\beta$, the infection rate $\sigma$ and the recovery rate $\gamma$ are assumed to be constant are not able to recover the stagnation observed in the infected case count in Wuhan. This suggests that the classical epidemiological models need to be revisited by including a time varying term which encapsulates the strict public health quarantine and isolation measures implemented in Wuhan, China post $24$ January 2020. \newline
By approximating this time varying quarantine strength with a neural network, we train the governing system of augmented SIR differential equations based on a loss function term obtained from the infected and recovered case count data generated by the Chinese National Health Commission. By training this governing system, we are able to not only approximate the maximum plateau value seen in the infected case count $30$ days post the implementation of quarantine control, but also recover a monotonically increasing quarantine strength function; indicating the increasing severity of quarantine measures with time. At the peak of its quarantine measures, we predict that about $70\%$ of the infected population was effectively isolated and prevented from spreading infection to the non-infected population. From the model, we can also recover the effective reproduction number, $R(t)$ from the model which is seen to decrease from $R(t) \approx 1.5$ to $R(t) < 1$ a month post $24$ January 2020, indicating the efficacy of quarantine and isolation measures in curtailing the spread of CoVID in Wuhan. \newline
We then used an optimized neural network-augmented SIR model to forecast the quarantine control strength and effective reproduction number in Wuhan. Our model predicts a stagnation of the quarantine strength function, $Q(t)$ as China eases the strength of quarantine and isolation measures in Wuhan. Due to a stagnation of the quarantine strength function, a stagnation in the effective reproduction number at $R(t) <1$ is estimated from end of March 2020. However, our results also include a warning that, although the infected case count in Wuhan is stagnating, this does not imply that quarantine measures  in Wuhan should be relaxed immediately. Since quarantined and isolated individuals may not have completely recovered, relaxing the quarantine measures may lead to a relapse in the infection spread and a subsequent increase in the effective reproduction number to $R(t) >1$. Thus, it may be wise to not relax quarantine measures after sufficient time has elapsed, during which maximum of the quarantined/isolated people are recovered. 
\iffalse
This indicates that since quarantined and isolated individuals may not be recovered, relaxing the quarantine measures may lead to a relapse in the infection spread and a subsequent increase in the effective reproduction number to $R(t) >1$. Thus, it may be wise to not relax quarantine measures after sufficient time has elapsed, during which maximum of the quarantined/isolated people are recovered. 
\fi
Finally, it should be noted that our model has a number of limitations. We do not consider the cases when the quarantined/isolated population come into contact with the non-infected population and lead to the transmission of disease. Although it has become evident that CoVID transmission also occurs through asymptomatic individuals, we do not make a distinction between symptomatic and asymptomatic individuals in our model. Irrespective of these assumptions, we believe this to be a first study to quantify the effective of quarantine measures implemented in Wuhan, with an interpretable physical model aided by machine learning techniques, involving very few free parameters.
\section*{Materials and Methods}
\textbf{Model 1: Without quarantine control} \newline
For modelling the number of CoVID cases without quarantine control in Wuhan, China, we use the classic SEIR epidemiological model, which has been employed in a number of prior studies, such as for modelling the SARS outbreak \cite{SEIR1, SEIR2, SEIR3} as well as the CoVID outbreak \cite{read2020novel, tang2020estimation, wu2020nowcasting}. In this compartment model, the entire population is divided into four compartments: the susceptible $S$; the exposed $E$; the infected $I$; and the recovered population, $R$. As the disease spreads, population redistribution occurs within these four compartments so that the total population $N = S + E + I +R$ remains constant at any given time $t$. In this model, we assume that there is (a) homogeneous mixing (b) identical disease process for every individual (c) equal susceptibility for each individual (d) timescale of the epidemic is much faster than the birth and death rate and (e) no spreading occurs through animals. Irrespective of these assumptions, the basic SEIR model serves as an interpretable model, often shedding light on the processes involved in shaping the disease transmission; making it a useful quantitative tool. The SEIR model is governed by the following system of nonlinear ordinary differential equations \citep{SEIR3, wang2020phase}

\begin{equation*}
    \cfrac{dS}{dt} = -\beta S I /N
\end{equation*}
\begin{equation*}
    \cfrac{dE}{dt} = \beta S I /N - \sigma E
\end{equation*}
\begin{equation*}
    \cfrac{dI}{dt} = \sigma E - \gamma I
\end{equation*}
\begin{equation*}
    \cfrac{dR}{dt} =  \gamma I
\end{equation*}

In addition to the SEIR model, the SIR model is also implemented to estimate the number of CoVID cases that would have resulted without quarantine control in Wuhan, China. In the SIR model, the population is instead divided into three compartments: susceptible $S$; infected $I$; and recovered $R$. Making similar assumptions as the SEIR model, the SIR model is governed by a lower-order system of ordinary differential equations according to 

\begin{equation*}
    \cfrac{dS}{dt} = -\beta S I /N
\end{equation*}
\begin{equation*}
    \cfrac{dI}{dt} = \beta S I /N - \gamma I
\end{equation*}
\begin{equation*}
    \cfrac{dR}{dt} =  \gamma I
\end{equation*}

In both the above models, $\beta$ is the transmission rate, $\gamma$ is the recovery rate. in the SEIR model, $\sigma$ is the infection rate. All these paraameters are constant in time.\newline
The reproduction number $R(t)$ in the SEIR and SIR models is defined as
\begin{equation*}
    R(t) = \cfrac{\beta}{\gamma}
\end{equation*}

\subsection{Initial conditions - 24 Jan 2020}
The number of susceptible individuals was assumed to be equal to the population of Wuhan, $ S_{0} = 11$ million. Wuhan's own infected population was obtained from the data released by the Chinese National Health Commission and was found to be $I_{0} = 500$. The exposed population was assumed to be $20 \times I_{0}$ in accordance with \citep{read2020novel, wang2020phase} and the number of recovered individuals was set very close to $0$ as $R_{0} \approx 10$.  \newline
\subsection{Parameter estimation}
In both the above models, the time resolved data for the infected, $I_{\textrm{data}}$ and recovered, $R_{\textrm{data}}$  case count for CoVID-19 was obtained from Chinese National Health Commission. The optimal values of the parameters $\beta, \sigma, \gamma$ were obtained by performing a local sensitivity analysis \citep{cao2003adjoint, DBLP:journals/corr/abs-1902-02376} of the ODE problems in the SEIR and the SIR models described above, by minimizing the mean square error loss function ($L (\beta, \sigma, \gamma)$) defined as
\begin{equation}\label{loss}
    L (\beta, \sigma, \gamma) = ||\textrm{log}(I(t)) - \textrm{log}(I_{\textrm{data}(t)}||^{2} + ||\textrm{log}(R(t)) - \textrm{log}(R_{\textrm{data}(t)}||^{2} 
\end{equation}

The plots in figure 1 are obtained with the optimal parameters $\beta, \sigma, \gamma$ for the SEIR model and $\beta, \gamma$ for the SIR, obtained by minimizing the loss function described in (\ref{loss}). \newline \newline
\textbf{Model 2: With quarantine control} \newline
To study the effect of quarantine control in Wuhan, we consider the SIR epidemiological model for analysis. We chose this model to have as few free modelling parameters in the system, so that the physics of the infected and recovered case count evolution is not obscured through the use of a large number of parameters. To include quarantine control in the modelling, we augment the SIR model with a time varying term which depends on the number of infected cases ($Q(t) I$) where $Q(t)$ is the quarantine strength function and $I(t)$ is the infected case count. Recently, it has showed that neural networks can be used as function approximators to recover unknown entities of a system of equations \cite{Rackauckas20}. Thus, in general, $Q(t)$ can be represented by a $n$ layer deep neural network with weights $W_{1}, W_{2} \ldots W_{n}$, activation function denoted by $\sigma$ and the input vector $U = (S(t), I(t), R(t), T(t))$ as
\begin{equation}\label{NN-1}
   Q(t) = \sigma \left( W_{n} \sigma\left( W_{n-1} \ldots \sigma \left( W_{1} U\right)\right)\right) = NN(W, U)
\end{equation}

In this study, we choose a two layer densely connected neural network with $10$ weights in each layer and the $\textrm{relu}$ activation function.
Thus, the governing differential equation for the augmented SIR model is 

\begin{equation*}
    \cfrac{dS}{dt} = -\beta S I /N
\end{equation*}
\begin{equation*}
    \cfrac{dI}{dt} = \beta S I /N - \gamma I - Q(t) I = \beta S I /N - \gamma I - \textbf{NN(W, U)}I
\end{equation*}
\begin{equation*}
    \cfrac{dR}{dt} =  \gamma I
\end{equation*}
\begin{equation*}
    \cfrac{dT}{dt} =  Q(t) I = \textbf{NN(W, U)}I
\end{equation*}

where $T(t)$ is the number of quarantined individuals in the population, assumed to be very small at the start of the model, {\it i.e.} $T_{0} = 10$  and $NN(W, U)$ is the neural network used for approximating the quarantine strength $Q(t)$. The reproduction number $R(t)$ in the augmented SIR model is defined as
\begin{equation*}
    R(t) = \cfrac{\beta}{\gamma + Q(t)}
\end{equation*}
The augmented SIR ODE system is trained by minimizing the mean square error loss function ($L (W, \beta, \sigma, \gamma)$) defined as
\begin{equation}\label{loss2}
    L (W, \beta, \sigma, \gamma) = ||\textrm{log}(I(t)) - \textrm{log}(I_{\textrm{data}(t)}||^{2} + ||\textrm{log}(R(t)) - \textrm{log}(R_{\textrm{data}(t)}||^{2} 
\end{equation}

where $I_{\textrm{data}(t)}$ is the infected case count and $R_{\textrm{data}(t)}$ is the recovered case count. \newline
The plots in figure 2 are obtained with the optimal weights $W$ of the neural network and the optimal parameters $\beta, \gamma$ for the SIR model, obtained by minimizing the loss function described in (\ref{loss2}).\newline\newline
\noindent {\bf Conflicts of Interest}
The authors declare no conflicts of interest. \newline

\noindent {\bf Acknowledgment}
We are grateful to Haluk Akay, Hyungseok Kim and Wujie Wang for helpful discussions and suggestions.

\bibliography{Paper_Draft1}
\bibliographystyle{jfm}
% Note the spaces between the initials
\end{document}